\input{aipcheck.tex}
\documentclass[final]{aipproc}
\layoutstyle{8x11single}
\def\be{\begin{equation}}
\def\ee{\end{equation}}
\def\bea{\begin{eqnarray}}
\def\eea{\end{eqnarray}}
\def\ba{\begin{array}}
\def\ea{\end{array}}
\usepackage{bm}
\newcommand{\xbf}[1]{\mbox{\boldmath $ #1 $}}

\begin{document}
\title[]{Nucleon shape and electromagnetic form factors in the chiral constituent quark model}
\classification{13.40.Gp, 12.39.Fe} \keywords{Electromagnetic form
factors, chiral constituent quark model}
\author{Harleen Dahiya}{address={Department of Physics, Dr. B.R. Ambedkar National Institute of Technology,
Jalandhar, Punjab-144 011, India.}}
\author{Neetika Sharma}{address={Department of Physics, Dr. B.R. Ambedkar National Institute of Technology,
Jalandhar, Punjab-144 011, India.}}

\date{\today}

\begin{abstract}
The electromagnetic form factors are the most fundamental
quantities to describe the internal structure of the nucleon and
the shape of a spatially extended particle is determined by its
{\it intrinsic} quadrupole moment which can be related to the
charge radii. We have calculated the electromagnetic form factors,
nucleon charge radii and the intrinsic quadrupole moment of the
nucleon in the framework of chiral constituent quark model. The
results obtained are comparable to the latest experimental studies
and also show improvement over some theoretical interpretations.

\end{abstract}

\maketitle

\section{electromagnetic form factors}
The electromagnetic form factors are fundamental quantities of
theoretical and experimental interest to investigate the internal
structure of nucleon. An understanding of the nucleon form factors
is necessary to describe the strong interactions as they are
sensitive to the pion cloud and provide a test for the QCD
inspired effective field theories based on the chiral symmetry.
The knowledge of internal structure of nucleon in terms of quarks
and gluons degrees of freedom of QCD also provide a basis for
understanding more complex, strongly interacting matter. Further,
the measurements from deep inelastic scattering (DIS) experiments
have revealed a significant amount of strangeness quark content in
the nucleon \cite{emc} which has been further reinforced by the
experiments performed in the recent past. In particular, the
SAMPLE Collaboration has reported the proton strange form factor
as $G_M^s = +0.37 \pm 0.20 \pm 0.26 \pm 0.07 $ n.m. \cite{sample},
whereas the HAPPEX Collaboration has reported $G_M^s= +0.18 \pm
0.27$ n.m. \cite{happex}. Thus, the importance of electromagnetic
form factors increases further to determine the strange form
factors contribution in the nucleon.

The internal structure of nucleon is determined in terms of
electromagnetic Dirac and Pauli from factors $F_1(Q^2)$ and
$F_2(Q^2)$ or equivalently in terms of the electric and magnetic
Sachs form factors $G_E(Q^2)$ and $G_M(Q^2)$ \cite{sach}.
Experimentally, these are determined in elastic $ep$ scattering
cross section via the Rosenbluth separation \cite{rosen} as well
as using polarization transfer \cite{polarization}. The issue of
determination of the form factors has been revisited in the recent
past with several new experiments measuring the form factors with
precision at MAMI \cite{mami} and JLAB \cite{jlab}. It has been
shown that the proton form factors determined from the
measurements of polarization transfer \cite{jlab} were in
significant disagreement with those obtained from the Rosenbluth
separation \cite{rosen1}. This inconsistency leads to a large
uncertainty in our knowledge of the proton electromagnetic form
factors and urge the necessity for the new parameterizations and a
new analysis \cite{fit}. The measurement of neutron form factors
is even more difficult than that for the proton since the free
neutron target does not exist. They are usually extracted from the
measurement of electron-deuteron scattering or electron-helium
scattering.

The most general form of the hadronic current for a spin
$\frac{1}{2}$-nucleon with internal structure is given as \be
\langle B| J^\mu_{{\rm had}}(0)|B \rangle = \bar{u}(p')
\left(\gamma^\mu F_1(Q^2)+ i{\sigma^{\mu\nu}\over 2M}q_\nu
F_2(Q^2)\right) u(p), \label{ff} \ee where $u(p)$ and $u(p')$ are
the 4-spinors of the nucleon in the initial and final states
respectively. The Dirac and Pauli form factors $F_1(Q^2)$ and
$F_2(Q^2)$ are the only two form factors allowed by relativistic
invariance. These form factors are normalized in such a way that
at $Q^2=0$, they reduces to electric charge and the anomalous
magnetic moment in units of the elementary charge and the nuclear
magneton $\mu_N$, for example, \bea F^p_1(0)= 1\,,\,~~~~ F^p_2(0)
= \kappa_p = 1.793 \,, \,~~~~ F^n_1(0) = 0\,,\,~~~~ F^n_2(0) =
\kappa_n = -1.913\,.\eea

In analogy with the non-relativistic physics, we can associate the
form factors with the Fourier transforms of the charge and
magnetization densities. However, the charge distribution
$\rho(\bf{r})$ has to be calculated by a 3-dimensional Fourier
transform of the form factor as function of $\bf{q}$, whereas the
form factors are generally functions of $Q^2=\bf{q}{^2}-\omega^2$.
It would be important to mention here that there exists a special
Lorentz frame, the Breit or brick-wall frame, in which the energy
of the (space-like) virtual photon vanishes. This can be realized
by choosing $\bf{p}_1= -\frac{1}{2}\bf{q}$ and $\bf{p}_2=
+\frac{1}{2}\bf{q}$ leading to
$E_1=E_2 \,,$ 
$\omega=0$ and $Q^2=\bf{q}^2$. Thus, in the Breit frame, Eq.
(\ref{ff}) takes the following form \cite{sach} \be J_{\mu} =
\left( G_E(Q^2)\,, \iota
\frac{\bf{\sigma}\times\bf{q}}{2M}G_M(Q^2) \right)\,,
\label{eq}\ee where $G_E(Q^2)$ stands for the time-like component
of $J_{\mu}$ hence identified with the Fourier transform of the
electric charge distribution, whereas $G_M(Q^2)$ is interpreted as
the Fourier transform of the magnetization density. The Sachs form
factors $G_E$ and $G_M$ can be related to the Dirac and Pauli form
factors as \be G_E(Q^2)= F_1(Q^2)- \tau F_2(Q^2)\ ,~~~~~~~~~~
G_M(Q^2)= F_1(Q^2)+ F_2(Q^2)\,,\label{sach} \ee where
$\tau=\frac{Q^2}{4 M_N^2}$ is a measure of relativistic effects.

The  Fourier transform of the Sachs form factors can be expressed as
\be G_E ({\bf{q}}^2)=\int\rho({ \bf r}) e^{i{\bf q}\cdot{\bf r}} d^3
{\bf r} ~~~= \int\rho({\bf r})d^3 {\bf r}- {\frac{\bf{q}^2}{6}}
\int\rho({\bf{r}}){\bf r}^2 d^3 {\bf r} +\ ...\ , \ee where the
first integral yields the total charge in units of $e$, i.e., 1 for
the proton and 0 for the neutron, and the second integral defines
the square of the electric root-mean-square (rms) radius, $\langle
r^2 \rangle_E$. The density $\rho(\bf{r})$ is not an observable but
only a mathematical construct in analogy with the classical charge
distribution.

\section{charge radii and Intrinsic quadrupole moment of the nucleon}

Elastic $ep$ scattering experiments apart from showing that the
proton has a finite size \cite{rosenfelder}, have also provided
detailed information on the radial variation of the charge and
magnetization densities. Charge radii contain fundamental
information about the internal structure of the baryons. The shape
of a spatially extended particle is determined by its {\it
intrinsic} quadrupole moment \cite{Hen01}, corresponding to the
charge quadrupole form factor $G_{C2}(q^2)$ at zero momentum
transfer. For the spin $\frac{1}{2}$ baryons,  the information on
their intrinsic quadrupole moments can be obtained from the
measurements of electric $(E2)$ and Coulomb $(C2)$ quadrupole
transitions to excited states \cite{Tia03,Ber03}. The {\it
intrinsic} quadrupole moment of a nucleus with respect to the body
frame of axis is defined as \be Q_0=\int d^3r \rho({\bf r}) (3 z^2
- r^2)\,. \ee If the charge density is concentrated along the
$z$-direction (symmetry axis of the particle), the term
proportional to $3z^2$ dominates, $Q_0$ is positive, and the
particle is prolate shaped. If the charge density is concentrated
in the equatorial plane perpendicular to $z$ axis, the term
proportional to $r^2$ prevails, $Q_0$ is negative and the particle
is oblate shaped.

In order to obtain information on these observables, we use a
general parametrization (GP) method developed by the Morpurgo {\it
et al.} \cite{morp}. It has been shown that it is possible to
parameterize several hadronic properties using the general
features of quantum chromodynamics (QCD) using the GP method. The
most general form of the charge radius operator for the sum of
one-, two-, and three-quark terms can be expressed as
\be\label{rad} {r}^2 = A \sum_{i=1}^3 e_i {\bf 1} + B\sum_{i \ne
j}^3 e_i \, {\xbf{\sigma}_i} \cdot \xbf{\sigma}_j  + C\sum_{i \ne
j \ne k }^3 e_k \, \xbf{\sigma}_i \cdot \xbf{\sigma}_j\,,\ee where
$e_i$ and $\xbf{\sigma}_i$ are the charge and spin of the i-th
quark. The constants $A$, $B$, and $C$ can be determined from the
experimental observations on charge radius and quadrupole moments
and the baryon charge radii for the octet and decuplet baryons can
then be calculated by evaluating matrix elements of the operator
in Eq. (\ref{rad}) between three-quark spin-flavor wave functions
$|B \rangle $ as $\langle B |r^2| B\rangle$.

Similarly, the charge quadrupole operator composed of a two- and
three-body term in spin-flavor space as \bea \label{quad} { Q} & =
& B'\sum_{i \ne j}^3 e_i \left( 3 \sigma_{i \, z} \sigma_{ j \,z}
- \xbf{\sigma}_i \cdot \xbf{\sigma}_j \right) + C'\!\!\sum_{i \ne
j \ne k }^3 e_k \left( 3 \sigma_{i \, z} \sigma_{ j \, z} -
\xbf{\sigma}_i \cdot \xbf{\sigma}_j \right). \eea Baryon decuplet
quadrupole moments $ Q_{B^*}$ and octet-decuplet transition
quadrupole moments $ Q_{B \to B^*}$ are obtained by calculating
the matrix elements of the quadrupole operator in Eq. (\ref{quad})
between the three-quark spin-flavor wave functions $\vert B
\rangle $ as \bea \label{matrixelements} Q_{B^*} = \left \langle
{B^*} \vert { Q}{\vert B^*} \right \rangle \,, ~~~~~~~~~~ Q_{B \to
B^*} = \left \langle {B^*} \vert { Q} \vert B \right \rangle
\,,\eea where $B$ denotes a spin $\frac{1}{2}$ octet baryon and
$B^*$ a spin $\frac{3}{2}$ decuplet baryon.

Calculations have been carried out for the charge radii and
intrinsic quadrupole moments in the quark model and pion cloud
model \cite{Hen01}, which lead to the several interesting
observations. It is found that the intrinsic quadrupole moment of
the proton is given by the negative of the neutron charge radius
and therefore positive, whereas the intrinsic quadrupole moment of
the $\Delta^+$ is negative. This corresponds to a prolate proton
and an oblate $\Delta^+$ deformation.

\section{chiral constituent quark model}

One of the most successful model in the non-perturbative regime of
QCD is the chiral constituent quark model with configuration
mixing ($\chi$CQM$_{{\rm config}}$)\cite{hd}.  The basic process
in the $\chi$CQM is the emission of a GB by a constituent quark
which further splits into a $q \bar q$ pair as $ q_{\pm}
\rightarrow {\rm GB}^{0} + q^{'}_{\mp} \rightarrow (q \bar q^{'})
+q_{\mp}^{'}\,, \label{basic} $ where $q \bar q^{'} +q^{'}$
constitute the ``quark sea'' \cite{cheng,song,johan}. The
effective Lagrangian describing interaction between quarks and a
nonet of GBs is ${\cal L} = g_8 \bar q \Phi q\,,$ \bea q =\left(
\ba{c} u \\ d\\ s \ea \right),& ~~~~~& \Phi = \left( \ba{ccc}
\frac{\pi^o}{\sqrt 2} +\beta\frac{\eta}{\sqrt
6}+\zeta\frac{\eta^{'}}{\sqrt 3} & \pi^+ & \alpha K^+   \\ \pi^- &
-\frac{\pi^o}{\sqrt 2} +\beta \frac{\eta}{\sqrt 6} +
\zeta\frac{\eta^{'}}{\sqrt 3}  &  \alpha K^o
\\ \alpha K^-  & \alpha \bar{K}^o & -\beta \frac{2\eta}{\sqrt 6}
+\zeta\frac{\eta^{'}}{\sqrt 3} \ea \right), \eea where $g_8$ and
$\zeta$ are the coupling constants for the singlet and octet GBs.
SU(3) symmetry breaking is introduced by considering $M_s >
M_{u,d}$ as well as by considering the masses of GBs to be
nondegenerate $(M_{K,\eta} > M_{\pi}$ and $M_{\eta^{'}}
> M_{K,\eta})$ {\cite{hd,cheng,song,johan}. The parameter
$a(=|g_8|^2$) denotes the transition probability of chiral
fluctuation of the splittings $u(d) \rightarrow d(u) +
\pi^{+(-)}$, whereas $\alpha^2 a$, $\beta^2 a$ and $\zeta^2 a$
respectively denote the probabilities of transitions of~ $u(d)
\rightarrow s  + K^{-(o)}$, $u(d,s) \rightarrow u(d,s) + \eta$,
 and $u(d,s) \rightarrow u(d,s) + \eta^{'}$.

The spin structure of a baryon can be defined as $\hat B \equiv
\langle B|{\cal N}|B \rangle\,,$ where $|B\rangle$ is the baryon
wave function  and $\cal N$ is the number operator $ {{\cal
N}}=n_{u_{+}}u_{+} + n_{u_{-}}u_{-} + n_{d_{+}}d_{+} +
n_{d_{-}}d_{-} + n_{s_{+}}s_{+} + n_{s_{-}}s_{-}\,,$ the
coefficients of the $q_{\pm}$ giving the number of $q_{\pm}$
quarks.

The wave function for the octet baryons with spin-spin generated
configuration mixing can be expressed as \be |B\rangle \equiv
\left|8, {\frac{1}{2}}^+ \right \rangle = \cos \phi \frac{1}{\sqrt
2}(\chi^{'} \varphi^{'} + \chi^{''} \varphi^{''}) \psi^{s}(0^+) +
\sin \phi\frac{1}{2}[(\varphi^{'} \chi^{''} +\varphi^{''}\chi^{'})
\psi^{'}(0^+) + (\varphi^{'} \chi^{'} -\varphi^{''}
\chi^{''})\psi^{''}(0^+)] \,. \label{mixed} \ee For the decuplet
baryons, we have \bea |B^{*} \rangle \equiv |10, {\frac{3}{2}}^+
\rangle = \chi^{s} \varphi^s \psi^s(0^+) \,,\eea where $\chi$,
$\varphi$ and $\psi$ are the spin, isospin and spatial
wavefunctions \cite{yaoubook}.

\section{Results and Discussion}

In the non relativistic limit the electric and magnetic form factor
of the baryons can be expressed as \bea G_E(Q^2) &=& \langle B|
\sum^3_{j=1} e_i e^{-\iota Q.r_i}| B \rangle\,,~~~~~~ G_M(Q^2) =
\langle B|\sum^3_{j=1} \mu_i\sigma_{iz} e^{-\iota Q.r_i}|
B\rangle\,. \eea For the case of proton and neutron, we have \bea
G_E^p(Q^2)=\cos^2 \phi \langle \psi ^s| e^{-\iota Q.r_3}|\psi ^s
\rangle + \frac{\sin^2 \phi}{2}\left(\langle \psi^{\lambda}|
e^{-\iota Q.r_3}|\psi^{\lambda} \rangle + \langle \psi^{\rho}|
e^{-\iota Q.r_3}|\psi ^{\rho} \rangle \right)+ \sqrt{2} \sin \phi
\cos \phi \langle \psi^s|e^{-\iota Q.r_3}|\psi ^{\lambda} \rangle\,,
\eea \be G_E^n(Q^2) = -\frac{\sin \phi \cos \phi}{\sqrt{2}} \langle
\psi ^{s}|e^{-\iota Q.r_3}|\psi ^{\lambda} \rangle\,, \ee
\bea G_M^p(Q^2)&=&\bigg[{\cos^2 \phi} \left(\frac{4}{3}\mu_u-
\frac{1}{3}\mu_d \right) \langle \psi ^s|e^{-\iota Q.r_3}|\psi^s
\rangle + \frac{\sin \phi \cos \phi}{\sqrt 2}
\left(\frac{10}{3}\mu_u+ \frac{2}{3}\mu_d \right) \langle \psi
^s|e^{-\iota Q.r_3}|\psi ^{\lambda} \rangle \nonumber
\\ &+&\frac{\sin^2 \phi}{4} \left( \left(2\mu_d \right) \langle
\psi^{\lambda}|e^{-\iota Q.r_3}| \psi^{\lambda} \rangle+
\left(\frac{8}{3} \mu_u - \frac{2}{3} \mu_d \right) \langle
\psi^{\rho}|e^{-\iota Q.r_3}|\psi ^{\rho} \rangle \right) \bigg]\,,
\eea \bea G_M^n(Q^2)&=&\bigg[ {\cos^2 \phi}\left(-\frac{1}{3}\mu_u+
\frac{4}{3}\mu_d \right) \langle \psi ^s|e^{-\iota Q.r_3}|\psi^s
\rangle + \frac{\sin \phi \cos \phi}{\sqrt 2}\left(\frac{2}{3}\mu_u+
\frac{10}{3}\mu_d
\right)\langle \psi^s|e^{-\iota Q.r_3}|\psi ^{\lambda} \rangle \nonumber \\
&+&\frac{\sin^2 \phi}{4} \left( \left(2\mu_u \right) \langle
\psi^{\lambda}|e^{-\iota Q.r_3}| \psi^{\lambda} \rangle+
\left(-\frac{2}{3}\mu_u+ \frac{8}{3}\mu_d \right) \langle
\psi^{\rho}|e^{-\iota Q.r_3}|\psi ^{\rho} \rangle \right) \bigg]\,.
\eea

Using the GP method, the charge radii for the spin $\frac{1}{2}$
ground state baryons is expressed as \be r^2= (A-3B)\sum_i e_i +
3(B-C)\sum_i e_i \sigma_{i z}\,, \ee where $\sum_i e_i=1$, $\sum_i
e_i \sigma_{i z}= 1$ for proton and $\sum_i e_i =0$, $\sum_i e_i
\sigma_{i z} =-\frac{2}{3}$ for neutron.

In $\chi$CQM, the proton and neutron charge radii can be expressed
as \be r^2_p= (A- 3B)(1- a -2 a \alpha^2)+ 3(B- C)\left( \cos^2
\phi \left(1 -\frac{a}{3} (4+ 2\alpha^2+ \beta^2+ 2\zeta^2)\right)
+ \sin^2 \phi \left( \frac{1}{3} -\frac{a}{9}(6+ \beta^2+
2\zeta^2) \right) \right)\,, \ee \be r^2_n= (A-3B)(1- a \alpha^2)+
3(B- C)\left( \cos^2 \phi \left(-\frac{2}{3}+ \frac{a}{9} (3 +
9\alpha^2+ 2\beta^2+ 4\zeta^2)\right) - \sin^2 \phi
\frac{a}{3}\left( 1- \alpha^2 \right) \right)\,.\ee We find that
the $N \to \Delta$ quadrupole moment is related to the neutron
charge radius as $Q_{p \to \Delta^+} = Q_{n \to \Delta^0} =
\frac{1}{\sqrt{2}}\, r_n^2$ which is experimentally well satisfied
and also show improvement over some theoretical interpretations.
Thus, the neutron charge radius plays an important role in the
hadron dynamics. It sets the scale not only for the charge radius
splitting within and between flavor multiplets but also for the
size of quadrupole moments and the corresponding intrinsic baryon
deformation. The results can further be substantiated by a
measurement of the baryon charge radius and other transition
quadrupole moments.

\section{ACKNOWLEDGMENTS}
H.D. would like to thank the organizers of HADRON09 and Department
of Science and Technology, Government of India for financial
support.

\end{document}